\documentclass[12pt]{iopart}

\usepackage{iopams} 
\usepackage{epsf}
\usepackage{graphicx} 
\begin{document}

\title{Pitfalls on the determination of the universality class of radial clusters}

\author{S. C. Ferreira Jr.$\dagger$\footnote[3]{To whom correspondence should be addressed (silviojr@ufv.br)} and S. G. Alves$\ddagger$}

\address{\small $^\dagger$Departamento de F\'{\i}sica, Universidade Federal Vi\c{c}osa, 36571-000, Vi\c{c}osa, MG, Brazil \\ $^\ddagger$Departamento de F\'{\i}sica, Universidade Federal de Minas Gerais, CP 702, 30161-970, Belo Horizonte, MG, Brazil}

\begin{abstract}
The self-affinity of growing systems with radial symmetry, from tumors to grain-grain displacement, has devoted increasing interest in the last decade. In this work, we analyzed features about the interface scaling of these clusters through large scale simulations (up to $3\times 10^7$ particles) of two-dimensional growth processes with special emphasis on the off-lattice Eden model. The central objective is to discuss an important pitfall associated to the evaluation of the growth exponent $\beta$ of these systems. We show that the $\beta$ value depends on the choice of the origin used to determine the interface width. We considered two strategies frequently used. When the width is evaluated in relation to the center of mass (CM) of the border, the exponent obtained for the Eden model was $\beta_{CM}=0.404\pm0.013$, in very good agreement with previous reported values. However, if the border CM is replaced by the initial seed position (a static origin), the exponent $\beta_0=0.333\pm 0.010$, in complete agreement with the KPZ value $\beta_{KPZ}=1/3$, was found. The difference between  $\beta_{CM}$ and $\beta_{0}$ was explained through the border CM fluctuations that grow faster than the overall interface fluctuations. Indeed, we show that the exponents $\beta_0$ and $\beta_{CM}$ characterize large and small wavelength fluctuations of the interface, respectively. These finds were also observed in three distinct lattice models, in which the lattice-imposed anisotropy is absent.

\end{abstract}


\submitto{\it J. Stat. Mech.}

\maketitle
\section{Introduction}

Interfaces in dynamic systems are present everywhere in nature, ranging from thin film deposition to biological growth. The scaling analysis of these interfaces constitutes a procedure widely used to characterize the underlying dynamics of these growth processes \cite{Barabasibook,Meakinbook, Vicsekbook}. One class of them that has attracted increasing interest along the past decade is the interface scaling of biological systems that exhibit radially symmetric growth \cite{Ferreira1998, Lacasta, Ferreira2002, Ferreira2003, Galeano, Bru1998, Bru2003, Bru2004, Buceta}. Initially, the scaling analysis of such biological systems was essentially restricted to the theoretical models \cite{Ferreira1998, Lacasta, Ferreira2002, Ferreira2003}. However, several biological experiments as, for instance, plant callus evolution \cite{Galeano} and the growth of malignant cells and tumoral explants \cite{Bru1998, Bru2003, Bru2004, Buceta}, have been reported in the past few years. Self-affine interfaces of radially symmetric patterns were also found in experiments of grain-grain displacement in Hele-Shaw cells in the quasi static regime \cite{Pinto}.

The Eden model \cite{Eden} is the simplest discrete example that generates radially symmetric patterns with self-affine interfaces. It was initially designed to describe biological pattern formation. The original version was studied on a square lattice, in which an occupied site represents a cell. The simulation begins with a single cell at the center of the lattice and the growth rules are as follows: at each step a site of the cluster periphery (an occupied site with at least one empty nearest neighbor)  is chosen at random and one of its empty nearest neighbors (NN) is selected with equal probability and occupied. Variants of the Eden rules were studied and the original model is commonly called Eden B \cite{Meakinbook}. From the biological viewpoint, the Eden model is unrealistic, but it produces interfaces with a rich scaling usually analyzed through the interface width $w$ defined as the root mean square deviation of the interface around its mean value
\begin{equation}
w=\left[\frac{1}{N}\sum_{i=1}^N\left(r_i-\bar{r}\right)^2\right]^{1/2},
\label{eq:w}
\end{equation}
where a set of $N$ distances $r_i$ represents the interface and $\bar{r}$ is the mean value of these distances. 

In the case of interfaces grown from a $d$ dimensional substrate of linear size $L$ \cite{Barabasibook}, the interface width commonly behaves as: $w\sim t^\beta$, for $t\ll L^z$, and $w\sim L^\alpha$, for $t\gg L^z$. The exponents $\beta$, $\alpha$ and $z$ (growth, roughness and dynamic exponents, respectively) are related by ${\alpha}=\beta{z}$. A given set of values of these exponents defines a universality class and can reveal fundamental properties of the interface dynamics. Examples in $1+1$ dimensions with exactly known exponents include the universality classes of Edwards-Wilkinson ($\beta=1/4$, $\alpha=1/2$, and $z=2$) \cite{EW}, Kardar-Parisi-Zhang ($\beta=1/3$, $\alpha=1/2$, and $z=3/2$) \cite{KPZ}, and Mullins-Herring ($\beta=3/8$, $\alpha=3/2$, and $z=4$) \cite{MH}, also known as EW, KPZ, and MH universality classes, respectively. In general, a universality class is related to a dominant physical process of the interface dynamics \cite{Barabasibook}. The EW, KPZ and MH universality classes are related to local relaxation, lateral growth, and surface diffusion, respectively. This interface analysis was applied to the growth of several types of tumors suggests that this biological growth dynamics belongs to the MH universality class \cite{Bru2003}. These experiments, which have a very significant impact because they reveal a universal dynamics of tumors, were grounded on a interface scaling analysis using Eq. (\ref{eq:w}) for radial tumors. However, as we will discuss along this paper, there is an important pitfall associated to this procedure that can lead to erroneous conclusions about the growth exponent $\beta$ and, consequently, to the universality class of the process.

Simulations of Eden clusters grown from a flat substrate show that the model belongs to the KPZ universality class \cite{Vicsekbook, Kertesz, Devillard}. However, the shape of the Eden clusters grown from a seed is very sensitive to the lattice anisotropy \cite{Zabolitzky,Batchelor}. Zabolitzky and Stauffer \cite{Zabolitzky} simulated clusters of the Eden A model\footnote[1]{In this version, the empty sites neighboring the cluster interface are chosen at random and then occupied} with $N\simeq 10^9$ particles on square lattices and observed a complex behavior of the interface width. For small clusters, a relatively good agreement with KPZ growth exponent ($\beta\approx 1/3$) was observed, contrasting with the linear dependence on time ($\beta\rightarrow 1$) obtained for asymptotically large clusters. The value $\beta=1$ is due to the diamond-like shape of the cluster imposed by the square lattice anisotropy \cite{Batchelor1998}. In order to determine the growth exponent of round Eden clusters, off-lattice simulations with  $N\approx2\times 10^5$ particles were done by Wang \textit{et al.} \cite{Wang,Wang2} by considering the center of mass (CM) of the cluster borders as the origin for the evaluation of the roughness. They measured an exponent $\beta=0.396$ and claimed that this value is close to the KPZ exponent.

In this work, we demonstrate through large scale simulations ($N>3\times 10^7$) that the border CM fluctuations in the off-lattice Eden model are not negligible and the random motion of the border CM determines the exponent value observed by Wang \textit{et al.}. Indeed, we found a growth exponent $\beta_0=0.333\pm 0.010$ very close to $1/3$ when an origin fixed on the initial seed is used. When the cluster CM (all cells are taken into account) is used, the exponent asymptotically converges to $\beta_0$. We also applied these ideas to three distinct on-lattice growth models in which the lattice-imposed anisotropy is absent.  The paper is organized as follows. In Sec. \ref{sec:model}, the simulation procedures for Eden model are described. In Sec. \ref{sec:result}, the results for Eden Model are presented and discussed while the on-lattice models and the scaling analysis are presented in the section \ref{app}.  Finally, some conclusions are drawn in Sec. \ref{sec:conclu}.

\section{\label{sec:model}Off-lattice Algorithm}

\begin{figure}[hbt]
\begin{center}
\includegraphics[width=8.5cm,height=!]{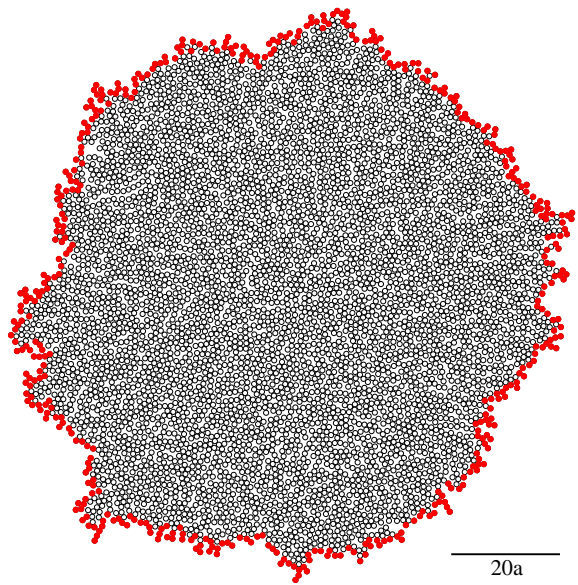}\\
\end{center}
\caption{\label{fig:pads} (color online) A small Eden cluster with $6000$ particles. The border is depicted in red.} 
\end{figure}

The simulations supporting the aforementioned results used the off-lattice algorithm proposed by Wang \textit{et al.} \cite{Wang}, in which the particles are represented by discs of diameter $a$ and the growth rules are the following.
\begin{enumerate}
\item An active cell (able to grow) is introduced on a plane.
\item A cell is selected randomly among the active ones. The intervals along which an adjacent cell can be grown without overlapping any existing cells are identified. Then, a new cell is grown in a direction randomly chosen in the allowed intervals.
\item If there are no possible growth directions, the cell is labeled as inactive.
\end{enumerate}
Also, since our interest is focused on the interface scaling, we introduced an optimization where any active cell inside a central core of radius $r_c$ is labeled as inactive. Since the inactivation of the particles near or belonging to the interface must be avoided, $r_c=0.8\bar{r}$ was chosen, where $\bar{r}$ is the mean radius of the interface. This optimization was used only for $\bar{r}>300a$. With these procedures, we grew clusters exceeding $3\times 10^7$ particles (more than two orders of magnitude larger than those obtained in previously reported simulations \cite{Wang}). In FIG. \ref{fig:pads}, a typical growth pattern and the corresponding border\footnote[2]{The border is defined as the set of cells that forms an external layer impenetrable for incoming cells. Consequently, the spaces between consecutive border cells is smaller than a cell diameter.} are illustrated. The mean density of cells inside the clusters has the closely constant value $\rho=0.633\pm0.001$, slightly lower than the density estimated by Wang \textit{et al.} ($\rho=0.6500\pm 0.0008$) \cite{Wang}. This difference is due to the divisions occurring inside the region $r<r_c$ in the original Wang algorithm which are forbidden in our modified rules. 

In order to evaluate the interface width, four methods were used to define the distances in Eq. (\ref{eq:w}). In the first one, the border CM method, $r_i$ is the distance from the instantaneous border CM to a site on the cluster border. In the second one, the cluster CM method, the CM of the entire cluster is used instead of the border one. In the third one, the seed method, the distance from the CM is replaced by the distance from the initial seed (a static origin). Finally, in the last one, the sector method, the border is divided in $k$ sectors of equal angular separation and $w$ is defined as the average of the standard deviation of the distances from the seed along each sector. Since the dynamics of the model occurs essentially on the border, we can do an analogy with the Eden model grown from flat substrates \cite{Vicsekbook} and consider the time proportional to the number of peripheral particles or, equivalently, to the mean radius.

\section{\label{sec:result} Off-lattice simulations}

\begin{figure}[hbt]
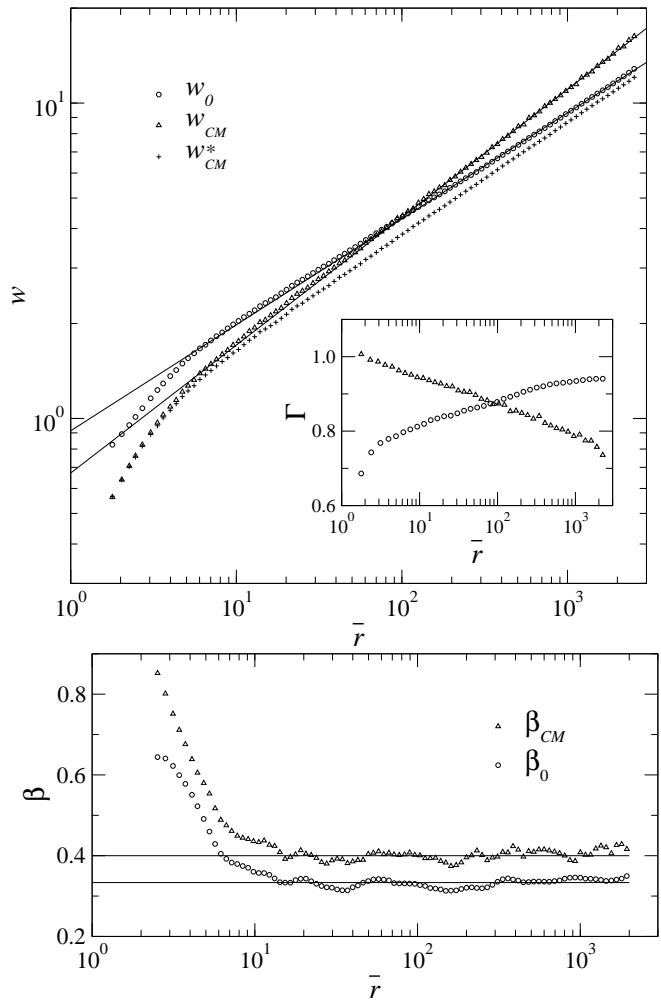

\includegraphics[width=8.5cm,height=!]{ws.eps} \\
\vspace*{0.5cm}
\includegraphics[width=8.5cm,height=!,clip=true]{beta.eps}
\caption{\label{fig:rugo} In the top, interface width evaluated using the border CM ($w_{CM}$), the cluster CM ($w^*_{CM}$), and the seed ($w_{0}$) methods is shown. The straight lines correspond to the power law fits for $\bar{r}\ge10^2a$. The inset shows the ratios $\Gamma=w^*_{CM}/w_{CM}$ (triangles) and $\Gamma=w^*_{CM}/w_{0}$ (circles). In the bottom, the local slope (the local growth exponent) is plotted as a function of the system size. The horizontal lines represent the slopes $1/3$ and $2/5$. All these curves result from $10^3$ independent samples. $\bar{r}$ and $w$ are given in cell diameter unities.}
\end{figure}

Figure \ref{fig:rugo} shows the interface width evaluated through the previously mentioned methods except the sector one. One can clearly observe distinct power laws for the roughness evaluated through the border CM and the seed methods. The corresponding exponents are $\beta_{CM}\approx 0.40$ and $\beta_0 \approx 0.33$, respectively.  The first value is in very good agreement with the simulations performed by Wang \textit{et al.} \cite{Wang} ($\beta=0.396$), whereas the second one is in excellent agreement with the KPZ universality class ($\beta_{KPZ}=1/3$). The last result confirms, for the first time, the claim that radial off-lattice Eden clusters belong to the KPZ universality class. The inset of FIG. \ref{fig:rugo}, in which the ratios between the interface width evaluated through distinct procedures are plotted, shows that the cluster CM and seed methods have the same asymptotic scaling ($\bar{r} > 10^3a$ or, equivalently,  $N>10^7$), while distinct growth exponents are observed in the intermediate intervals. Certainly, this transient can lead to wrong conclusions about the universality classes of experiments where the system can not grow forever \cite{Galeano,Bru1998,Bru2003,Bru2004}. We evaluated the local slope of the plots $\ln w$ against $\ln\bar{r}$ as shown in the bottom of FIG. \ref{fig:rugo}. As one can see, the exponents $\beta_0$ and $\beta_{CM}$ are clearly distinct and oscillate around the their expected values $1/3$ and $2/5$, respectively, for $\bar{r}\gtrsim 20 a$. Considering these fluctuations as estimates of the exponent uncertainties, we obtained $\beta_0=0.333\pm 0.010$ and $\beta_{CM}=0.404\pm 0.013$.

\begin{figure}[hbt]
\includegraphics[width=8.5cm,height=!]{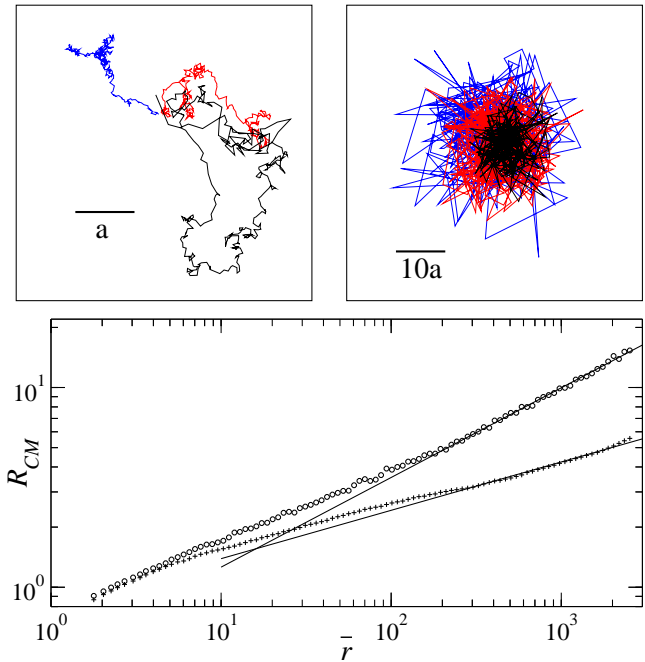}
\caption{\label{fig:cm} (color online) Typical trajectories of the cluster  (top left) and the border (top right) centers of mass. The colors correspond to 342 (black), 684 (red), and 1026  (blue) steps (definition in the text). In the bottom, the border (circles) and cluster (squares)  CM mean distances from the initial seed are plotted as functions of the mean radius. $R_{CM}$ and $\bar{r}$ are given in cell diameter unities.}
\end{figure}

The previous difference can be better understood by analyzing the CM evolution. Figure \ref{fig:cm} shows three stages of typical trajectories of the cluster and border CMs along a simulation (top). In these walks, a step is defined as the CM displacement when the cluster radius of gyration\footnote[4]{The definition of radius of gyration is $r_g=(\sum_{i=1}^{N}r_i^2/N)^{1/2}$, where $r_i$ represent the distances from the initial seed.} increases by a cell diameter. The difference between the trajectories is evident. In the former, the CM wanders through a region with a few cell diameters and follows a trajectory of low fractal dimension. In the later, the CM wanders around a region of increasing amplitude and magnitude with the same order than the interface width. Consequently, the trajectory of the border CM is more compact than that of the cluster CM. The bottom of FIG. \ref{fig:cm} shows the CM mean distance from the initial seed $R_{CM}$ as a function of the mean radius. As can be seen, for both cluster and border CMs, the mean distance grows approximately as a power law $R_{CM} \sim \bar{r}^\gamma$ for $\bar{r}>200a$. The exponents $\gamma_b=0.45\pm0.04$ and $\gamma_c=0.24\pm0.05$ were found for the border and cluster CMs, respectively. These results elucidate the differences between the growth exponents. When the cluster CM is used, its fluctuations around the initial seed grow slower than the interface fluctuations ($\beta>\gamma_c$). So, in the asymptotic limit the CM fluctuations are negligible and the growth exponent converges to $\beta_0$ (inset of FIG. \ref{fig:rugo}). In contrast, the border CM fluctuations increase faster than the interface fluctuations ($\gamma_b>1/3$) and, hence, they do not become asymptotically negligible. These findings might be easily verified in experimental essays such as those related to the callus growth \cite{Galeano}, tumor evolution \cite{Bru1998,Bru2003,Bru2004}, or grain-grain displacement in Hele-Shaw cells \cite{Pinto}.

\begin{figure}[ht]
\includegraphics[width=8.5cm,height=!]{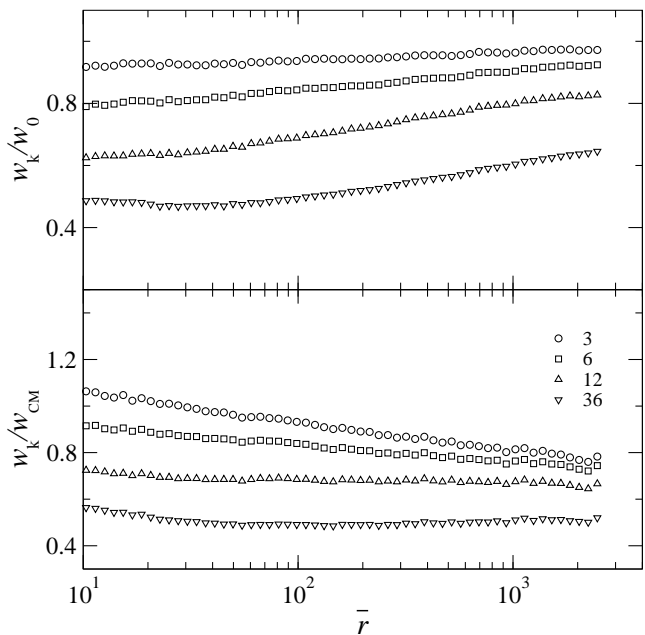}
\caption{\label{fig:wk} Ratios between interface widths determined through the sector ($w_k$), the seed ($w_0$), and  the border CM ($w_{CM}$) methods as  functions of the mean radius $\bar{r}$. The corresponding numbers of sectors are indicated in the legend and $\bar{r}$ is given in cell diameters unities.}
\end{figure}

In FIG. \ref{fig:wk}, the results for the sector method are confronted with those  obtained using the border CM and seed methods. This figure shows the ratios $w_k/w_0$ and $w_k/w_{CM}$ (definitions in the legend of FIG. \ref{fig:wk}) as functions of $\bar{r}$. A very peculiar behavior arises from these curves: if the interface is divided in a small number of sectors, the growth exponent quickly converges to $\beta_0$, whereas the scaling of border CM method is observed for a large number of sectors. For intermediate number of sectors ($k=12$), a crossover from $\beta_{CM}$ to $\beta_0$ can be perceived, suggesting that the growth exponent asymptotically reaches the value $1/3$. So, the KPZ universality class observed in the Eden clusters results from the large wavelength fluctuations of the interface. Also, one can infer that the border CM fluctuations are straightforwardly related to the small wavelength fluctuations of the border. This analysis may be relevant for the characterization of several experiments. For example, distinct scales for interface fluctuations have been recently identified in cell membrane of single macrophages during the  phagocytosis process \cite{Agero,Neto}

\section{\label{app} On-lattice models}

\begin{figure*}[hbt]
\includegraphics[width=14.5cm,height=!,clip=true]{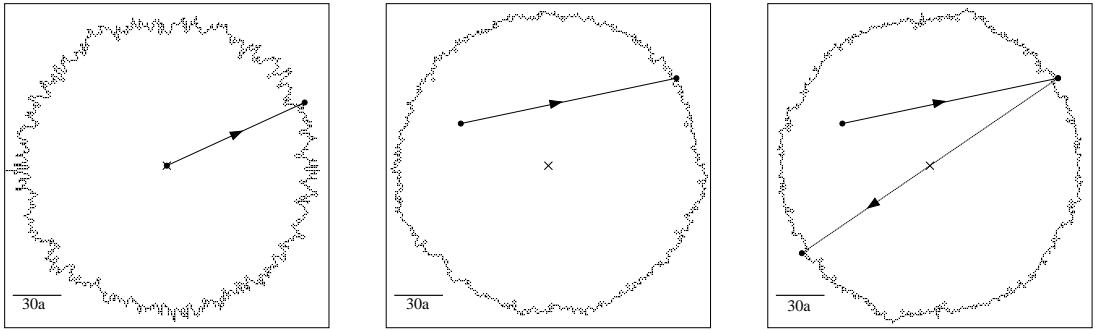}\\
\caption{\label{fig:models} Schematic representation of the on-lattice models. The clusters are representative samples of actual simulations. \textit{Model I} (left): the particle leaves the center towards the border; \textit{Model II} (middle): the particle follows a ballistic trajectory with random initial positions and directions; \textit{Model III} (right):  a first particle is grown through a random ballistic trajectory (continuous line) and a second one is added to the opposite side of the cluster (dashed line).}
\end{figure*}

In order to corroborate the scaling concepts based on the off-lattice Eden model simulations, we proposed three different growth rules. We chose on-lattice models due to the easiness for proposing and implementing new rules. However, these models should avoid the undesirable lattice anisotropy effects. The rules described in the sequence fulfill this requirement. In all models, the simulations begin with an occupied site at the center of the lattice and the growth rules at each time step are the following:
\begin{enumerate}
\item \textit{Model I}. A particle is released at the center of the lattice (on an occupied site) and follows a ballistic trajectory at a random direction while it does not reach an empty site (FIG. \ref{fig:models},left).
\item \textit{Model II}. Like in Model I, except that the particle is released at any occupied site chosen at random (FIG. \ref{fig:models},middle). 
\item \textit{Model III}. Firstly, the same rule of Model II is implemented. Secondly, other particle is added to the symmetric position in the cluster, passing through the center of the lattice, with probability $p$ (FIG. \ref{fig:models},right).  This model allows one to control the CM fluctuations. In particular, the CM is static for $p=1$.
\end{enumerate}

The isotropy of the patterns was confirmed using noise reduction methods (see \cite{Meakinbook} or \cite{Vicsekbook} for details about the method). The growth exponents are summarized in table \ref{tab:tab1}. As one can see, independently of the model, the exponent $\beta_{CM}$ has a value close to $2/5$ whereas the exponent $\beta_0$ for the model I is neatly different from the other ones. This results together with Eden model simulations suggest that $\beta_{CM}=2/5$ is a universal exponent.
\begin{table*}[ht]
\begin{center}
\begin{tabular}{ccc}
\hline\hline
Model & ~~~$\beta_0$~~~ & ~~~$\beta_{CM}$~~~ \\ \hline
Model I & $0.284 \pm 0.009$ & $0.38 \pm 0.02$ \\ 
Model II & $0.209\pm 0.006$ & $0.40\pm0.05$ \\ 
Model III ($p=0.75$) & $0.217\pm 0.007$ & $0.40\pm 0.04$\\
Model III ($p=0.90$) & ~~$0.213 \pm 0.009 $~~& $0.39\pm 0.04$ \\ \hline\hline

\end{tabular}
\end{center}
\caption{\label{tab:tab1} Growth exponents of the isotropic on-lattice growth models.}
\end{table*}

\begin{figure}[hbt]
\includegraphics[width=8.5cm,height=!,clip=true]{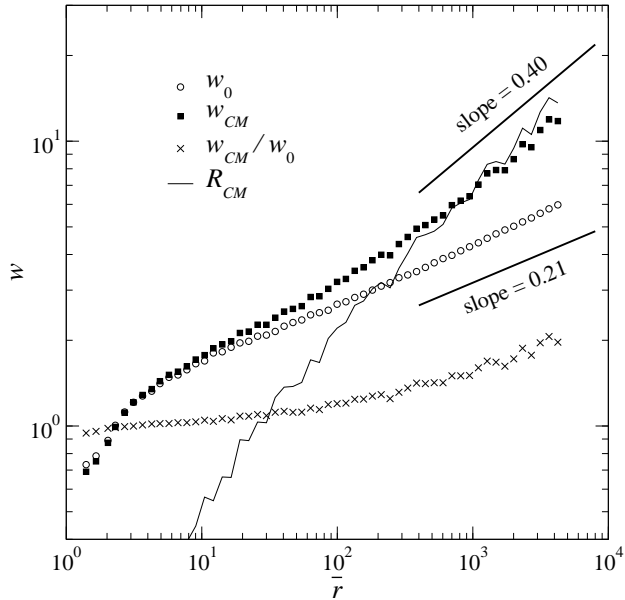}\\
\caption{\label{fig:basim} Interface width evolution for the model III with $p=0.90$. The averages were done over 140 independent samples.  $\bar{r}$ and $w$ are given in cell diameter unities.}
\end{figure}

Model III allows one to control the CM fluctuations and, consequently, to verify their roles on the growth exponents. Figure \ref{fig:basim} shows the quantities $w_0$, $w_{CM}$, $R_{CM}$ and the ratio $w_{CM}/w_0$ for $p=0.90$ as functions of time. These curves provide evidences of a crossover induced by the amplification of the CM fluctuations. Indeed, the scaling regime with $\beta_{CM}\ne \beta_0$ emerges when $R_{CM}\sim w_0$.

\section{\label{sec:conclu} Summary}

In conclusion, we drew new considerations about the interface scaling of round clusters using large scale simulations of isotropic radial clusters, mainly the off-lattice Eden model. This approach reveals a subtle pitfall that can be present in the analysis of experiments with radial symmetry. Indeed, we show that the growth exponent depends on the strategy adopted to measure the interface width. For the particular case of the off-lattice Eden model, the expected value $1/3$ is found when a fixed origin is used as reference. Otherwise, when the border CM is used as the origin, we found the growth exponent $\beta_{CM}=0.404\pm0.013$ very close to $2/5$, in very good agreement with previous reports \cite{Wang}. These differences arise from the border CM fluctuations increasing faster than those of the interface. We also show that the exponents $\beta_0$ and $\beta_{CM}$ are associated to the large and small wavelength fluctuations, respectively. These analyzes were corroborated by three distinct on-lattice models for which the lattice induced anisotropy is absent.

It is important to emphasize that the essential features presented in this work were obtained from a very simple approach (inclusion of the CM analysis) not previously considered, while the models were used only as a support to these analysis. Moreover, the difficulty for observing these features in the radial clusters previously studied, particularly the Eden model, lies on the lattice-induced cluster misshape that masks these fluctuations. Finally, it is also important to stress that the growth exponent has been used to draw conclusions about the underlying dynamics of systems with unquestionable scientific relevance (e.g. Refs. \cite{Bru1998, Bru2003, Bru2004}) without caution with the pitfalls presented in this work.

\bigskip

\textbf{Acknowledgments}\\ The authors thank to M. L. Martins, M. S. Couto, and I. L. Mensezes-Sobrinho for the critical reading and discussions about the manuscript. This work was partially supported by the Brazilian Agencies CNPq, CAPES, and FAPEMIG.

\end{document}